\documentclass[preprint,superscriptaddress,showpacs]{revtex4}

\usepackage{graphicx}

\begin{document}

\title{Coherent light transport in a cold Strontium cloud}
\author{Y. Bidel}
\affiliation{Laboratoire Ondes et D\'esordre, FRE 2302, 1361 route des Lucioles
F-06560 Valbonne, France}
\author{B. Klappauf}
\affiliation{Now at: Optoelectronics Research Center, University of
Southampton, SO17 1BJ Southampton, United Kingdom}
\author{J.C. Bernard}
\affiliation{Laboratoire Ondes et D\'esordre, FRE 2302, 1361 route des Lucioles
F-06560 Valbonne, France}
\author{D. Delande }
\affiliation{Laboratoire Kastler Brossel, Universit\'e Pierre et Marie Curie,
F-75252 Paris, France}
\author{G. Labeyrie}
\author{C. Miniatura}
\author{D. Wilkowski}
\email{wilkowsk@inln.cnrs.fr}
\homepage{http://www-lod.inln.cnrs.fr/}
\author{R. Kaiser}
\affiliation{Laboratoire Ondes et D\'esordre, FRE 2302, 1361 route des Lucioles
F-06560 Valbonne, France}

\date{\today{}}

\begin{abstract}

We study light coherent transport in the weak localization regime using
magneto-optically cooled strontium atoms. The coherent backscattering cone is
measured in the four polarization channels using light resonant with a ${\it
J_g=0}\rightarrow{\it J_e=1}$ transition of the Strontium atom. We find an
enhancement factor close to 2 in the helicity preserving channel, in agreement
with theoretical predictions. This observation confirms the effect of internal
structure as the key mechanism for the contrast reduction observed with an
Rubidium cold cloud (see: Labeyrie et al., PRL \textbf{83}, 5266 (1999)).
Experimental results are in good agreement with Monte-Carlo simulations taking
into account geometry effects.

\end{abstract}


\maketitle

During the past twenty years, the outstanding development of mesoscopic
physics led to a critical inspection of coherent effects in wave transport.
 First motivated by electronic transport in conducting devices
\cite{electrons}, the underlying physical ingredients proved to be relevant to
any linear waves and in particular to light. This triggered active research in
the field of optics during the past two decades \cite{Kuzmin} leading to the
observation of coherent backscattering \cite{Wolf} and universal conductance
fluctuations \cite{Scheffold} to quote a few. A challenge in this field is
still the observation of strong localization of visible light. It was recently
reported for near-infrared light using semi-conductors powders \cite{Wiersma},
but the interpretation of the experiment in term of Anderson localization was
questioned \cite{scheffold99}. Cold atoms have been quite recently considered
as promising scattering media to achieve strong localization
\cite{Nieuwenhuizen}. Indeed, they constitute perfectly monodisperse samples of
resonant point-dipole scatterers with large cross-sections. Moreover high
spatial density is achieved by adequate trapping techniques \cite{bec,katori}.

In this letter we report the observation of coherent backscattering (CBS) of
light on cold strontium atoms in the weak localization regime $kl \gg 1$ ($k$
is the light wavenumber and $l$ the elastic mean free path). CBS is an
interferential enhancement of the \emph{average} scattered intensity reflected
off a disordered scattering medium \cite{qqchose}. It originates from a
two-wave constructive interference (near exact backscattering) between waves
travelling along a given scattering path and its reversed counterpart. For
classical scatterers, bearing on general symmetry arguments valid in the
absence of any magnetic field, the CBS interfering amplitudes have been shown
to have equal weights at exact backscattering in the so-called parallel
polarization channels \cite{Bart}. In the $lin\| lin$ channel the incoming and
detected light fields have same linear polarization. In the $h\| h$ channel,
both light fields are circularly polarized with the same helicity, that is
opposite polarizations (because the CBS signal is emitted in the backward
direction). In the perpendicular channels, nothing ensures the equality of the
two interfering amplitudes and the contrast of the interference is decreased.
The single scattering events require a separate treatment as the direct and
reversed paths coincide and do not contribute to the CBS enhancement in the
backward direction. For spherically symmetric scatterers, single scattering
does not contribute in the $lin \perp lin$ and $h \| h$ channels. Thus, the CBS
contrast (peak to background ratio) is predicted and has been observed to be
exactly 2 in the helicity preserving polarization channel $h \| h$
\cite{Wiersma2}. Using an atomic gas at resonance, a dynamic breakdown of the
CBS effect can occur due to the scatterers motion during the transit time of a
photon inside the medium. This restricts the RMS velocity $\delta v$ below a
critical velocity given by $v_c= \Gamma /k$ (where $\Gamma$ is the width of the
atomic dipole resonance), a condition which is well fulfilled for a laser
cooled atomic gas \cite{labeyrie9900}. The quantum internal structure of atoms
has also severe consequences for coherent light transport in atomic media. A
degeneracy in the groundstate induces a dramatic scrambling of the CBS effect
\cite{jonckmuller}. This has been first experimentally observed with a cold
Rubidium sample on a $J_{g}=3 \rightarrow J_{e}=4$ transition
\cite{labeyrie9900}. These results highly motivated the use of nondegenerate
groundstate atoms, like strontium, to benefit from full interference effects in
coherent transport.

The cold strontium (Sr) cloud is produced in a magneto-optical trap (MOT). The
transverse velocity of an effusive atomic beam, extracted from a $500^{\circ}$C
oven, is immediately compressed with a 2D optical molasse. A 27~cm long Zeeman
slower then reduces the longitudinal velocity to within the capture velocity
range of the MOT ($\sim$ 50~m/s). The Zeeman slower, molasses, MOT, and probe
laser beams at 461~nm are generated from the same frequency-doubled source.
Briefly, a single-mode grating stabilized diode laser and a tapered amplifier
are used in a master-slave configuration to produce 500~mW at 922~nm. The
infrared light is then frequency doubled in a semi-monolithic standing wave
cavity with an intra-cavity KNbO$_\textrm{3}$ crystal. The cavity is resonant
for the infrared light while the second harmonic exits through a dichroic
mirror providing 150~mW of tunable single-mode light, which is then frequency
locked on the 461~nm $^{1}$S$_{0}$-$^{1}$P$_{1}$ strontium line in a heat pipe.
We use acousto-optic modulators for subsequent amplitude and frequency
variations. The MOT is made of six independent trapping beams of
5.2~mW/cm$^{2}$ each, red-detuned by $\delta=-\Gamma$ from the resonance. The
saturation intensity is 42.5 mW/cm$^{2}$ and the natural width of the
transition is $\Gamma /2\pi=32$~MHz. Two anti-Helmoltz coils create a 100~G/cm
magnetic field gradient to trap the atoms. A small population loss to
metastable states is repumped to the ground state using two additional red
lasers. The best achieved optical thickness of our Sr MOT is $b\approx3$. It is
deduced from transmission measurements of a resonant probe through the cloud
shortly after switching the MOT off. Note that because the optical thickness of
the atomic cloud is larger than one, the imaging of the cloud does not yield a
signal proportional to the atomic density (flattening at the center) and the
whole process thus overestimates the size of the cloud (see discussion below).
The number of trapped atoms $N \simeq 10^{7}$ is derived from the MOT
fluorescence signal. From a CCD image the RMS radius of the cloud has been
estimated at 0.65 mm yielding a mean free path $l\approx0.5$ mm ($kl \simeq
7000$). The RMS velocity of the atoms is less that 1 m/s, well below the
critical velocity $v_c=$ 15 m/s.

The detailed experimental procedure for the CBS observation has been published
elsewhere \cite{labeyrie9900}. For the Sr experiment, the signal is obtained
using a collimated resonant probe beam with a waist of 3~mm. To avoid any
effects linked to the saturation of the optical transition (non-linearities,
inelastic radiation spectrum) \cite{nonlinear}, the probe intensity is weak
(saturation parameter $s=0.02$). The scattered light is collected in the
backward direction by placing a CCD camera in the focal plane of an achromatic
doublet. The angular resolution of our apparatus is 0.1~mrad, roughly twice the
CCD pixel angular resolution. To avoid recording the MOT fluorescence signal
while recording the CBS signal, a time-sequenced experiment is developed. The
trapping beams and the magnetic field gradient are switched off during the CBS
acquisition sequence (duration $100~\mu s$) and then switched on to recapture
the atoms (duration 95\% of the 6~ms total cycle time). This procedure  also
eliminates any possible unwanted nonlinear wave mixing processes. The whole
time sequence is then repeated as long as necessary for a good signal-to-noise
ratio (typically 15 minutes in the experiment). During the CBS sequence, the
image field is opened (and then closed during the MOT sequence) thanks to a
mechanical chopper. During the CBS probe interaction time, each atom scatters
about 200~photons on average but always remains in resonance since the mean
atomic velocity increase is far below $v_c$. Consequently, most of them are
recaptured during the following MOT sequence. The CBS images (see
Fig.\ref{image}) are finally obtained by subtracting the background image taken
without cold atoms. This background image is recorded in the absence of the
magnetic gradient during all the acquisition time. We thus checked that the
fluorescence signal from the residual Sr atoms was negligible.

In the helicity preserving channel ($h\| h$), the enhancement factor is found
to be $\alpha=1.86\pm0.10$ with an optical thickness of $b=2.9$ (see
Fig.~\ref{coupe}), slightly lower the theoretical prediction $\alpha=2$.
Several experimental issues can explain the difference. First, the finite
angular resolution of the detection apparatus lowers the CBS enhancement factor
by an amount evaluated to $\delta\alpha\approx0.06$. Because single scattering
contributes more than 90\% of the total signal in the two authorized channels
(see Table~\ref{result}), the reduction of the cone contrast due to imperfect
polarization channel isolation in the $h\| h$ is not negligible. We have
measured, in the limit of low optical thickness where single scattering
dominates over multiple scattering, the fraction of detected light in the
forbidden $h \| h$ channel with respect to the total scattered light. We found
a channel isolation about $5.10^{-4}$ leading to $\delta\alpha\approx0.03$.
Note that single scattering depolarization induced by stray magnetic field acts
here like an imperfect polarization isolation. For this reason, its impact on
the cone reduction has been minimized during the channel isolation procedure.
Another possible source of contrast reduction is a Faraday effect induced by
the residual magnetic field \cite{lenke00}. It turns out that, despite the huge
Verdet constant in the atomic gas medium \cite{labeyrie01}, its effect should
be smaller than the previously discussed ones. We also checked that the finite
transverse size of the laser beam has no significant influence on the signal.
Taking into account the systematic errors, we find that the CBS enhancement
factor should rather be $\alpha=$1.91, consistent with the measured value. A
remaining (but yet uncontrolled) source of error in determining $\alpha$ is
certainly an imperfect estimation of the background level measured at angles
large compared to the cone width $\theta \gg \Delta\theta_{\mathrm{CBS}}$.

In the other polarization channels, we observe lower enhancement factors as
predicted by the theory (see table \ref{result}). In the $lin \| lin$ and $h
\perp h$ channels, the small enhancement factors are mainly due to the strong
single scattering contribution -- see the relative incoherent background values
given in table \ref{result} -- which is very important since the optical
thickness is not very large. In the $ lin \perp lin$ channel (where single
scattering is absent), the relatively high contrast value is explained by the
low optical thickness. Indeed, in this situation, short scattering paths
dominate and double scattering is known to exhibit full interference contrast
in all polarization channels \cite{albada87}. In Table \ref{result}, we also
show data obtained with a Monte-Carlo (MC) calculation, where the amplitude of
a multiple scattering path is computed as a function of the initial and final
polarizations and of the geometrical positions of the various scatterers. We
use a Gaussian distribution for the spatial density of scatterers and take into
account the spatial variations of the mean free path during the photon
propagation. Our numerical method is tantamount to computing the integral
involved in the configuration average using a MC procedure. Given a spatial
configuration of the scatterers, we compute simultaneously the various
scattering contributions at different scattering orders using the "partial
photon" trick \cite{partial_photon}. Typically, it is enough to launch less
than 1 million photons on the medium to get a good signal/noise ratio for the
CBS peak. For all polarization channels, there is a good agreement for the cone
height between experiments and MC simulations adjusted to take into account the
polarization channel isolation and angular resolution effects.

The experimental values $\Delta\theta_{\mathrm{CBS}}$ of the FWHM CBS angular
cone width are systematically higher (by a factor 1.4) that the ones given by
the MC simulation using the measured optical thickness $b$ and size of the
atomic cloud. As discussed above, our experimental procedure slightly
overestimates the size of the cloud. Modifying the size of the cloud (keeping
$b$ constant) results only in a global multiplication of the angular scale,
keeping identical both the enhancement factor and the cone shape. We are thus
inclined to think that the actual RMS radius of the cloud is 0.45 mm instead of
0.65 mm.  With this corrected value, we observe an excellent agreement between
MC and experimental data {\em in all polarization channels} (see
Fig.~\ref{coupe} and Table~\ref{result}). The angular dependence of the cone
shape in the linear channels reflects the anisotropy of the scatterer's pattern
\cite{albada87}. In the $lin \| lin$ channel, an elliptical shape with major
axis parallel to the incident polarization is predicted and indeed observed
(Fig. \ref{image}c). In the $lin\perp lin$ channel, the directions of maximum
scattering are tilted at $45^{\circ}$ from the incident polarization, yielding
a "clover-leafed" CBS cone shape (Fig. \ref{image}d).

To summarize, we measured the coherent backscattering cone in four different
characteristic polarization channels. Our results are in good agreement with a
Monte-Carlo calculation. The restoration of a full interference contrast in
coherent multiple scattering with atomic gases (as exemplified by the maximum
enhancement factor of 2 obtained in the helicity preserving channel) has
interesting potentialities for wave localization experiments with cold atoms.
For example, in the quest for Anderson localization (which could be obtained
only at high density where $kl \approx 1$) where interferences play a crucial
role, a $J_{g}=0 \to J_{e}=1$ transition appears to be a good choice, since a
degenerate internal structure is known to scramble the interference
\cite{jonckmuller}. A maximum enhancement factor of 1.2 was found in Rb
experiment \cite{labeyrie9900}. Is it now possible to increase the cloud
density to reach the Anderson localization threshold? For this purpose, cooling
strontium with the intercombination line in a dipole trap appears to be a
promising technique \cite{katori}.

The authors thank the CNRS and the PACA region for their financial support.
Laboratoire Kastler Brossel is laboratoire de l'Universit{\'e} Pierre et Marie
Curie et de l'Ecole Normale Sup{\'e}rieure, UMR 8552 du CNRS.

----------------------------------------------------------------
\begin{figure}[p]
\begin{center}
\end{center}
\caption{Isocontours of the coherent backscattering cones obtained on a cloud
on cold Strontium atoms in the four polarization channels. We plot the CBS
signal after substraction of the incoherent background, as a function of the
backscattering angle. All cone heights have been scaled to 1. For better
signal/noise the image have been symmetrized. The total angular range is 1
mrad. The lowest isocontour correspond to roughly 20\% of the peak intensity.
For the linear channels, the incident polarization is horizontal.}
\label{image}
\end{figure}

\begin{figure}[p]

\begin{center}
\end{center}
\caption{Angular scan of the CBS cone for the $h \| h$ polarization channel.
The optical thickness is $b=2.9$. The experimental data are represented with
open circles. For better signal/noise we perform an angular averaging of the
original image. The solid line is the result of a Monte-Carlo calculation
taking into account the geometry of the atomic cloud (Gaussian distribution of
the atomic density with
 variance 0.45 mm)
and experimental imperfections like
the polarization channel isolation and angular
resolution effects. The agreement is clearly excellent.
The measured width of the cone
is $\Delta\theta_{CBS}=0.50\pm0.04$
mrad.} \label{coupe}
\end{figure}

\begin{table}[p]
\caption {Comparison between the CBS enhancement factor and peak width measured
in the experiment with the results of a Monte-Carlo calculation, for optical
thickness $b=2$. In each polarization channel, the experimental enhancement
factor $\alpha$ is given with a $\pm 2\sigma$ error bar. For linear
polarization channels, the $\Delta\theta_{CBS}$ values are only given for scans
parallel to the incident polarization. The results of MC simulation (noted MC)
are given for a Gaussian distribution of the could with a variance 0.45 mm. The
experimental imperfections like the polarization channel isolation and angular
resolution effects have been taken into account in the MC simulation values
noted MC$^*$. The ``Background'' column show the relative contribution of the
channel of the total incoherent scattered intensity in the backward direction.}
\label{result}

\begin{tabular}[t]{|c|c|c|c|c|c|}
\hline Channel
        &
    & Background
        & $\alpha$
            & $\Delta\theta_{CBS}$ (mrad)\\
\hline
                &Exp.
                    & 7.5\%
                        & $1.77\pm0.13$
                            & $0.52\pm0.07$ \\
        ${\it h} \|{\it h}$
                & MC
                    & 7.8\%
                        & 2
                            & 0.48 \\

                &MC$^*$
                    & 7.8\%
                        & 1.87
                            & 0.52 \\
\hline
                &Exp.
                      & 92.5\%
                        & $1.17\pm0.03$
                            & $0.71\pm0.10$ \\
            ${\it h} \perp{\it h}$
                &MC
                    & 92.2\%
                        & 1.20
                            & 0.69 \\

                &MC$^*$
                    & 92.2\%
                        & 1.19
                            & 0.75  \\
\hline

                &Exp.
                    & 96.0\%
                        & $1.17\pm0.03$
                            & $0.9\pm0.2$\\
            ${\it lin} \|{\it lin }$
                &MC
                    & 95.5\%
                        & 1.24
                            & 0.92\\

                &MC$^*$
                    & 95.5\%
                        & 1.22
                            & 0.98\\
\hline
                &Exp.
                    & 4.0\%
                        & $1.59\pm0.20$
                            & $0.5\pm0.3$\\
        ${\it lin} \perp{\it lin }$
                &MC
                    & 4.5\%
                        & 1.74
                            & 0.48 \\

                &MC$^*$
                    & 4.5\%
                        & 1.62
                            & 0.49 \\
\hline
\end {tabular}

\end{table}

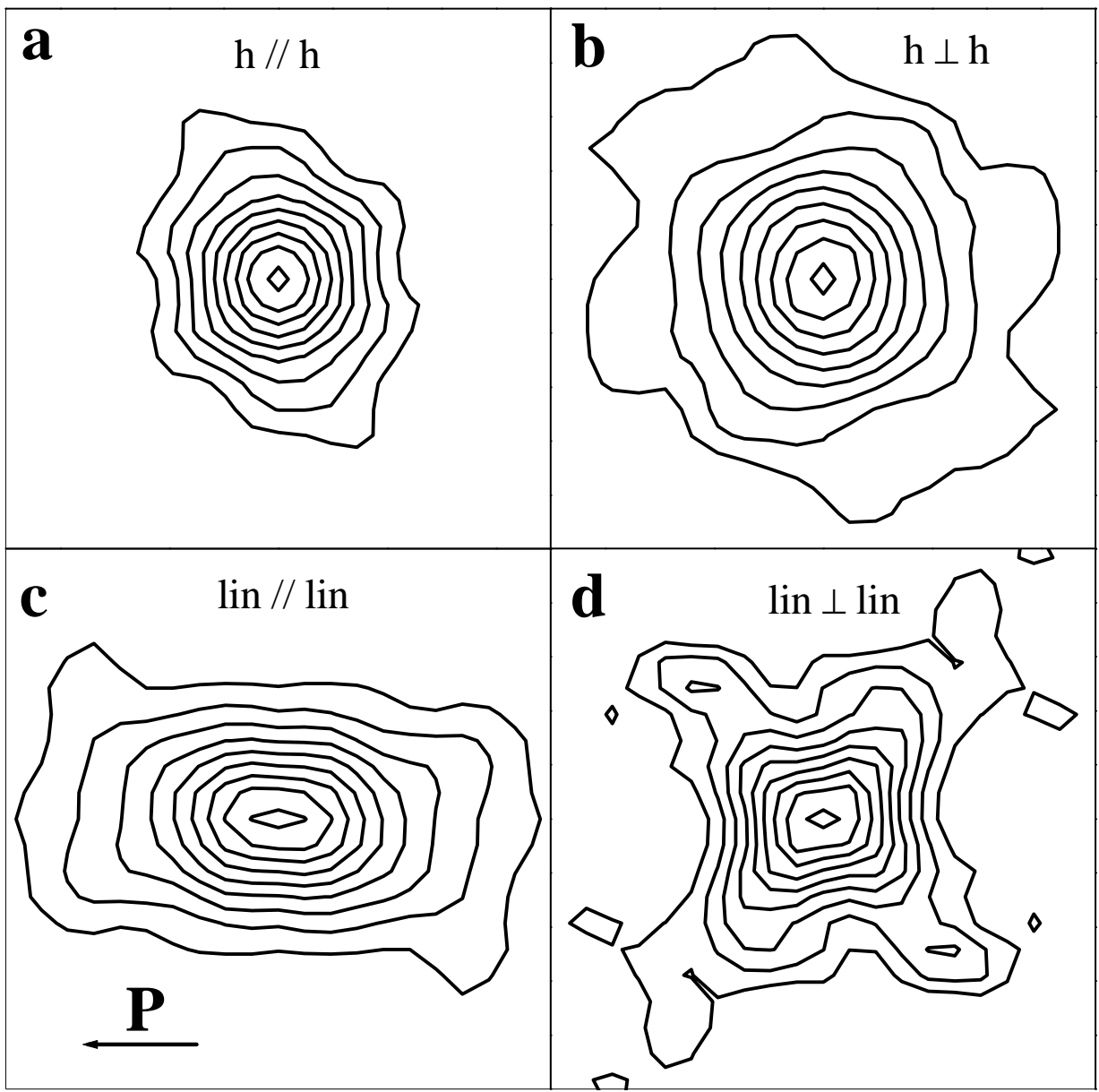
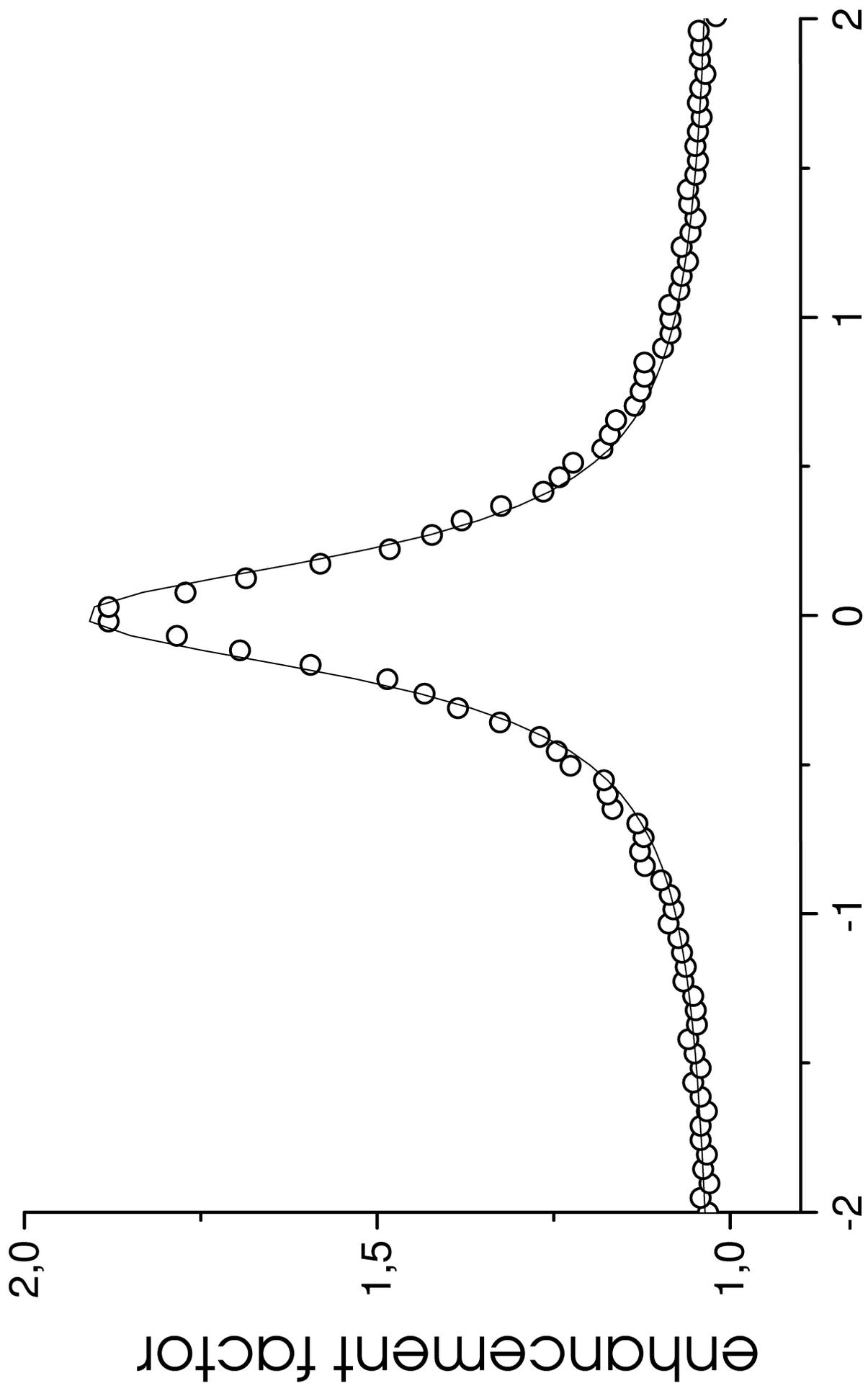
\end{document}